\documentclass{article}

\usepackage{graphicx}
\usepackage[pdftex]{hyperref}

\DeclareGraphicsExtensions{.eps,.jpg,.pdf}

\begin{document}
\bibliographystyle{plain}
\openup2.2\jot

{

\title{{\bf Efficient boundary integral solution for acoustic wave
scattering by irregular surfaces}}

\vskip 1.5 true cm
\author{Orsola Rath Spivack \& Mark Spivack}

\font\eightrm=cmr8


\maketitle

\vskip 1 true cm
{\eightrm
\begin{center}
Department of Applied Mathematics and Theoretical Physics,
The University of Cambridge CB3 0WA, UK

\end{center}
}

\vskip 1.5 true cm
\begin{abstract}
The left-right operator splitting method is studied for the efficient
calculation of acoustic fields scattered by arbitrary rough surfaces. 
Here the governing boundary integral is written as a sum of left- and
right-going components,  and the solution expressed as an
iterative series, expanding about the predominant direction of
propagation. 
Calculation of each term is computationally inexpensive both in time
and memory, and the field is often accurately captured using one or
two terms.    The convergence and accuracy are examined by comparison
with exact solution for smaller problems, and a series of much larger
problems are tackled.  The method is also immediately applicable to other
scatterers such as waveguides, of which examples are given.

\end{abstract}

\bigbreak

\vfill

}

\vskip 2 true cm

\bigbreak
 
\def\t{{\bf t}}

\def\x{{\bf x}}
\def\b{{\bf b}}
\def\H{{\bf H}}
\def\J{{\bf J}}
\def\Jinc{{\J_{inc}}}
\def\JJ{{\J_{0}}}
\def\JK{{\tilde\J_{inc}}}
\def\K{{\bf K}}
\def\L{{\bf L}}
\def\A{{\bf A}}
\def\C{{\bf C}}
\def\D{{\bf D}}
\def\n{{\bf n}}
\def\intf{\int_{-\infty}^{\infty}}
\def\intfs{\int_{{\rm surface}}}
\def\c{{\bf c}}
\def\d{{\bf d}}
\def\del{\partial}
\def\rbar{{\bf r}}
\def\E{{\bf E}}
\def\psiinc{{\psi_{inc}}}
\def\F{{\bf F}}
\def\L{{\bf L}}
\def\R{{\bf R}}
\def\u{{\bf u}}
\def\f{{\bf f}}
\def\g{{\bf g}}
\def\P{{\bf P}}
\def\p{{\bf p}}
\def\EQN{\eqno}
\def\beq{\begin{equation}}
\def\eeq{\end{equation}}

\vfil\break

\section {Introduction}
 
The calculation of acoustic scattering by extended rough 
surfaces remains a challenging problem both theoretically and
computationally (e.g. \cite{ogilvy1,vor1,warnick,saillard})
especially in the presence of strong multiple scattering.
This becomes acute at low grazing angles,   
where multiple scattering occurs for very slight roughness.
Boundary integral methods are flexible and often used for 
such problems but can be 
computationally intensive and scale badly with increasing wavenumber.
Much effort has therefore been devoted to this aspect, where possible
exploiting properties of the scattering regime.
For forward scattering in 2-dimensions, for example, provided 
roughness length-scales are large, the `parabolic integral equation
method' can be applied \cite{thorsos1,spivack2}.  
For electromagnetic problems, also formulated using boundary integrals,
the methods of ordered multiple interactions and left-right
splitting 
in both 2-d and 3-d (\cite{kapp1}-\cite{pino})
have been developed: here the scattered field is expressed as an
iterative series of terms of increasing orders of multiple scattering,
as described below. Approaches using conjugate gradient
solutions combined with fast multilevel multipole are also receiving
much attention.   An important exception which overcomes 
the dependence of computational expense
on wavenumber is \cite{chandler}, which has been applied to
surfaces with piecewise constant impedance data or scattering in 2-d
by convex polygons. 

\medbreak
A versatile recursive technique known as Multiple Sweep
Method of Moments was developed and analysed in \cite{colak,colak2} where it was
compared with Method of Ordered Multiple Interactions. This 
technique was shown to tackle `composite' problems for which the above method
diverges such as for a ship on a rough sea surface.
Other iterative solutions have been studied 
in \cite{macaskill1}.  
In addition theoretical results are available in various limiting regimes
(e.g. perturbation theory for small surface heights, $k\sigma \ll 1$
including periodic surfaces \cite{thorsos2,holford1,desanto1}, 
Kirchhoff approximation
\cite{meecham1,thorsos3}, or the small slope approximation
 \cite{vor3} which is accurate over a wider range of
 scattering angles than both of these). 
For arbitrary finite rough surfaces, however, 
validation is more difficult, and such results are therefore scarce.    

\medbreak
In this paper the Left-Right Splitting method is developed and applied to 
the problem of acoustic scattering in three dimensions 
by randomly rough surfaces.
For relatively small surfaces the results are validated by comparison
with numerical solution of the full boundary integral equation.
The principal aims are to validate the approach; to examine its
robustness and convergence as the angle of incidence changes;
and to consider further approximations which may reduce the
computation time.   The approach is applicable to a wide range of
interior and exterior scattering problems, and we give examples for
acoustic propagation in a varying duct, in addition to scattering from
large rough surfaces.

\medbreak
The mathematical principles of the method are the same as for the
two-dimensional problem \cite{spivack1} although implementation is
considerably more complicated:   The unknown field $\psi$ on the
surface is
expressed as the solution to the Helmholtz integral equation,
with the integration taken over the rough surface.  
This may be written formally as 
$A{\psi} = {\psiinc}$, 
where ${\psiinc}$ is
the incident field impinging (say) from the left, so that 
we require ${\psi}=A^{-1}{\psiinc}$.  
The region of integration is split into two, to the left
and right of the point of observation, allowing $A$ to be written
as the sum of `left' and `right' components,
say $(L+R){\psi} = \psiinc$. Roughly speaking $L$
represents surface interactions due to scattering 
from the left, and $R$ the residual scattering from the right.
The inverse of $A$ can formally be expressed as a series
\beq 
A^{-1}= L^{-1} - L^{-1} R L^{-1} + ...
\label{eq1} \eeq 
Discretization of the integral equation yields a block
matrix equation,  in which $L$ is the lower triangular part of 
the block matrix 
$A$ (including the diagonal) and $R$ is the upper triangular part.
Under the assumption that most energy is right-going, $L$ is the
dominant part of $A$, and the series can be truncated to
provide an approximation for ${\psi}$.   This approach has several
advantages. In terms of wavelength $\lambda$, 
evaluation of each term scales with the fourth rather than the 
sixth power of $\lambda$ required for $A^{-1}$;
subsequent terms (of which typically only the first one or two are needed)
have the same computational cost.  With further approximations this 
can be reduced to $\lambda^3$. However, this operation count is
only part of the story, because the low complexity and memory
requirement allow very large problems to be tackled without such
additional approximation.  
%
In addition the algorithm lends itself well to parallelisation, and
the speed scales approximately linearly with the number of processors.
\medbreak

\medbreak
In \S 2 the governing equations and left-right splitting approximation are
formulated.   The numerical details and main results are shown in \S 3.

\section{Formulation of equations}
 
\begin{figure}
\hskip 3 true cm
\includegraphics[height=6cm]{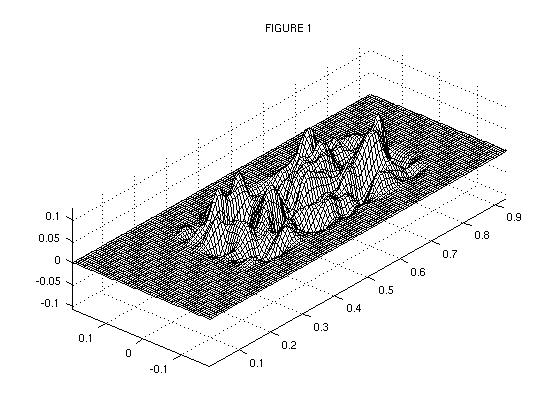} 
\caption{\label{example} Example rough surface}
\end{figure}

Consider a 3-dimensional medium with horizontal axes $x, y$ and vertical 
axis $z$ directed upwards, and let $k$ be the wavenumber.  
Let $S=s(x,y)$ be a 2-dimensional rough surface, 
varying about the plane $z=0$, which is
continuous and differentiable as a function of $x,y$ (see Figure
\ref{example}).
(Arbitrary scatterers can also be 
treated by the methods shown here; examples will be given later.)
Consider a time-harmonic acoustic wave $\psi$, obeying
the wave equation  $(\nabla^2+k^2)\psi=0$ in the region $z>s(x,y)$, 
resulting from  an incident wave $\psiinc$ at a small grazing angle
$\theta$ to  the horizontal plane. This may for example be a plane
wave or a finite beam. 
The axes can be chosen so that the principal direction of propagation 
is at a small angle to the $(x,z)$ plane. 
\vskip 1 true cm

We will treat the Neumann boundary condition, i.e. an acoustically hard
surface. The derivation for the Dirichlet condition is similar. Thus

\beq{\del \psi\over \del \n} = 0 \eeq
where $\n$ is the outward normal (i.e. directed out of the region $z>s(x,y)$).
The free space Green's function is given by 
\def\kbar{{\bf k}}
\beq G(\rbar,\rbar') = {e^{ik|\rbar-\rbar'|}\over 4\pi|\rbar-\rbar'|} .
\label{(5.1)} \eeq
The field at a point $\rbar$ in the medium is related to the surface
field by the boundary integral 
\beq \psi_{inc}(\rbar) = \psi(\rbar) - \int_S 
{\del G(\rbar,\rbar')\over \del n}
\psi (\rbar') d\rbar'
\label{eq4} \eeq
where $\rbar=(x,y,z)$ and $\rbar'=(x',y',s(x',y'))$, 
say, 
and taking the limit as $\rbar\rightarrow \rbar_s$ gives
\beq \psi_{inc}(\rbar_s) = \psi(\rbar_s) - \int_S 
{\del G(\rbar_s,\rbar')\over \del n}
\psi (\rbar') d\rbar'
\label{eq5} \eeq
where now $\rbar_s=(x,y,s(x,y))$.
The integrand is singular at the point $\rbar'=\rbar_s$, and
we must take care to interpret this integral as the limit 
of the integral in eq. (\ref{eq4}) as $\rbar\rightarrow\rbar_s$.

\medbreak
In order to treat the equation numerically it is convenient 
to write the integration  with respect to $x$,$y$, so that
eq. (\ref{eq5}) becomes 

\beq \psi_{inc}(\rbar_s) = \psi(\rbar_s) -  \int_{-\infty}^\infty 
\int_{-\infty}^\infty {\del G(\rbar_s,\rbar')\over \del n}
\psi (\rbar') ~\gamma(\rbar') dx' dy'
\label{eq6} \eeq
where (with very slight abuse of notation)
\beq
\gamma(\rbar') = \sqrt{1+
\left({\del s\over\del x'}\right)^2 + \left({\del s\over\del y'}\right)^2 }.
\label{(5.5)} 
\eeq
and the expression under the square root is evaluated at $\rbar'$.

\bigbreak
\subsection{Formal solution and splitting series}

The method of solution is analogous to that applied to the
electromagnetic problem in 2-d or 3-d \cite{spivack1,spivack4}.
The governing integral equation (\ref{eq6}) is expressed 
in terms of right- and left-going operators $L$ and $R$
with respect to the $x$-direction:

\beq \psi_{inc} (\rbar_s) = A \psi \equiv (L+R) \psi \label{eq8} \eeq
where $L$ and $R$ are defined (for an $L^2$ function $f$) by 
\begin{eqnarray}
Lf(\rbar) &=& f - \int_{-\infty}^\infty
\int_{-\infty}^{x} {\del G(\rbar,\rbar')\over\del n} f(\rbar') 
~\gamma(\rbar') ~dx' ~dy',
\\
~~~~~ Rf(\rbar) &=& - \int_{-\infty}^\infty \int_{x}^\infty 
{\del G(\rbar,\rbar')\over\del n} f(\rbar') ~\gamma (\rbar') ~dx' ~dy' 
\label{(5.7)} 
\end{eqnarray}
and $\rbar=(x,y,z)$, $\rbar'=(x',y',s(x',y'))$.  [For notational
conveneince $L$ is interpreted to include the contribution from the
singularity arising in (\ref{eq5}) when $\rbar \rightarrow \rbar'$.]

The region of integration is thus split into two with respect to $x$,
and 
the solution of equation (\ref{eq8}) can be expanded as a series,
given by 

\beq \psi = (L+R)^{-1} \psi_{inc} = 
 \left[L^{-1} - L^{-1} R L^{-1} + ... \right] \psi_{inc} . \label{series} \eeq
 The key observation is that at fairly low grazing angles the
 effect of $\R$ is in some sense small, so that the series converges quickly
 and can be truncated.    Define the $n$-th order approximation
as 
\beq
\psi_n=\sum_1^n ~ L^{-1} \left(R L^{-1}\right)^{n-1} ~ \psiinc .
\eeq
[Note that $L$ and $R$ depend on surface geometry and
wavenumber only, not on incident field; and that one might expect
convergence of the series (\ref{series}) for given $\psiinc$
but not uniform (norm) convergence of the series (\ref{eq1}).]
This corresponds physically to an assumption that surface-surface interactions
 are dominated by those `from the left', as expected in this scattering regime.
$L$ is large compared with $R$ 
first, because $L$ includes the dominant `diagonal' value; 
second because a predominantly right-going wave gives rise to more rapid
phase-variation in the integrand in $R$ than in $L$. 
(Although this depends on surface geometry and cannot in general be
quantified precisely, it 
occurs because in (\ref{eq5}) the phase in the Green's function kernel
decreases as the observation point is approached from the left and
 then increases to the right; whereas the phase of $\psi$  
tends to increase throughout, like that of the incident field.)
This is borne out 
numerically, with many cases of interest well-described using only one or
 two terms of the series.

\begin{figure}
  \includegraphics[height=3cm]{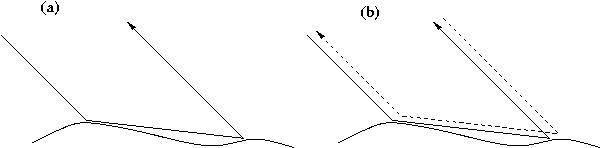} 
\caption{\label{schematic} 
Possible paths (a) at 1st iteration, and (b) at 2nd
iteration when reversible paths can occur and add coherently.}  
\end{figure}

The scattered field due to a given approximation $\psi_n$ 
is obtained by substitution back into the
boundary integral (\ref{eq4}). 
It is helpful 
to consider the significance of successive approximations
to this field in the ray-theoretic limit:
The first iteration contains ray paths which, before leaving the
 surface,  may have interacted with
the surface arbitrarily many times but only in a forward direction. 
The second includes most paths which have
changed direction twice: once via the operator $R$ and again via
$L^{-1}$; and so on (see Figure \ref{schematic}).   Thus 
the first iteration accounts for multiple scattering but not {\it
 reversible} paths which can 
occur when incident and backscatter direction are opposite;
these paths occur in pairs of equal length and therefore add
 coherently, giving rise to a peak in the
 backscattered direction (enhanced backscatter eg \cite{maradudin,macaskill2}) 
in strongly scattering regimes. 
We would therefore expect this to show initially at the second approximation.
\vskip 1 true cm

Having obtained this series, numerical treatment by surface
discretization is straightforward.  
(Discretization can equivalently be carried out before the
series expansion,  but it is more 
convenient, and analytically more transparent, to expand the
integral operator first.) 


\section{Numerical solution and results}

Although use of the series (\ref{series}) is
motivated by physical considerations and its terms provide
a convenient theoretical interpretation, the immediate advantage is 
computational:
if the surface is discretized using a rectangular grid of 
$M$ by $N$ points, with $M$ transverse steps ($y$ direction) and $N$ in
range ($x$), then $(L+R)$ becomes an $(MN\times MN)$ matrix, and 
exact inversion would take $O((MN)^3)$ operations. 
On the other hand evaluation of each 
term of eq. (\ref{series}) involves inversion of an 
$M\times M$ matrix at each of $N$ range steps, 
requiring  just $O(NM^3)$ operations and far
less memory. Assuming a resolution of say 10 points per
wavelength, this scales with $\lambda^4$.
There is an additional `matrix filling' component;
this also increases with $\lambda^4$, and in practice this 
is the dominant computational cost in the left-right splitting
algorithm (typically more than 90\% when $M\cong N$).

\subsection{Numerical solution}

The numerical treatment will now be outlined.
The notation $L$, $R$ will be used to refer
to the discretized forms of the integral operators where 
no confusion arises, and we will
focus on solution of the first term of (\ref{series}), i.e. inversion of $L$.
Although not evaluated explicitly as such, 
the matrix $L$ is conveniently viewed as an $N\times N$
lower-triangular block matrix 
whose entries are $M\times M$ matrices.  The system
can therefore be inverted by Gaussian elimination and back-substitution.
This is an $N$-step `marching' process, in which each diagonal 
$M\times M$ block is inverted in turn, corresponding to marching the
solution for the unknown surface field in the positive $x$ direction.
Choosing step-sizes $\Delta x$, $\Delta y$ 
we define
\begin{eqnarray} x &=&x_1,...,x_N, ~~~~~~x_n=n\Delta x \\
y &=&y_1,...,y_M, ~~~~~~y_m=m\Delta y  . 
\label{(5.9)} 
\end{eqnarray}
Denote the discretized surface values by 
\begin{eqnarray}  \psiinc(x_n,y_m) &= a_{nm} \\ 
\psi(x_n,y_m) &= b_{nm} ,
\end{eqnarray}
denote the area of each subintegration region by $\delta=\Delta x\Delta y$, 
and write 
$\sigma_{ij}=~\gamma(\rbar_{ij})$
where $\gamma$ (equation (\ref{(5.5)})) is evaluated at the point
$\rbar_{ij}=(x_i,y_j,s(x_i,y_j))$. 
This induces a discretization of (\ref{eq8}) and
at each point surface point $s(x_n,y_m)$ we get
\beq 
a_{nm} = \sum_{i=1}^N \sum_{j=1}^M ~A_{nmij}~ b_{ij} \label{(5.10)} 
\eeq
where 
\begin{eqnarray} 
A_{nmij} &=& \Delta x \Delta y ~\sigma_{ij}
\del G(\rbar_{nm},\rbar_{ij}) /  \del n 
~~~~~~~~~~~~~~(n\neq i, ~{\rm or}~ m\neq i) \\
A_{nmnm} &=& - \left[{1\over 2} - {\delta\over\sigma_{ij}^2}  
(s_{xx} +s_{yy})\right] 
\label{(5.11)} 
\end{eqnarray}
and again $\rbar_{ij}=(x_i,y_j,s(x_i,y_j))$.
For each value of $n$ this gives a set of $M$ equations.
%
%
%
%
Retaining just the first term in the iterative series (\ref{series}),
\beq
\psi \cong L^{-1} \psiinc , \label{(5.12)} 
\eeq
yields a set of equations identical
to (\ref{(5.10)}) except that the sum over $i$ has upper limit $n$:
\beq a_{nm} = \sum_{i=1}^n \sum_{j=1}^M ~A_{nmij}~ b_{ij} .
\label{(5.13)} 
\eeq
This is equivalent to integration over the half plane to the left of 
the line of observation ($x'\leq x_n)$.  
Now at each range step $x_n$, assuming that we have obtained 
the values $b_{im}$ for $i < n$, equation (\ref{(5.13)}) can be
rearranged to give 
\beq  a_{nm} - \sum_{i=1}^{n-1} \sum_{j=1}^M ~A_{nmij}~ b_{ij} 
 = 
\sum_{j=1}^M  ~A_{nmnj}~ b_{nj} \label{(5.14)} \eeq 
for $m=1,...,M$.  Everything on the left-hand-side is 
known or has been found at previous steps.
For each $n$ this gives a matrix equation, which we rewrite for
convenience as
\beq  \c_{n} = B_{n} \b_{n} \label{(5.15)} \eeq 
where the subscript $n$ indicates dependence on $x_n$ and we have
written the vectors in bold.
Therefore, $\b_{n}$ denotes solution values $\psi(x_n,y)$
at the range step $x_n$, and $B_n$ is the
$M\times M$ matrix (the $n$-th term on the diagonal of $L$) with elements 
\beq  
\left(B_n \right)_{mj} = A_{nmnj} . 
\eeq 
We thus require 
\beq   
\b_{n} = B_{n}^{-1} \c_{n} 
\label{abc}
\eeq 
for each $n$.  We solve (\ref{abc}) in turn for $n=1,...,N$,  
using each result to redefine the left-hand-side of
eq. (\ref{(5.15)}) and thus find the surface
field as defined by (\ref{(5.12)}).   Subsequent terms in the series
(\ref{series}) are evaluated in exactly the same way, with the 
`driving' term $\psiinc$ replaced by $R$ times the result of the
previous evaluation.

\bigbreak


\bigbreak
\subsection{Computational results}

One of the main applications is to irregular or randomly rough
  surfaces (for example sea surfaces or terrain).
 Statistically stationary surfaces with Gaussian
  statistics (normally distributed heights) are easily generated
  computationally with any prescribed spatial
  autocorrelation function (a.c.f.) $\rho(\xi,\eta)$, where
\beq
\rho(\xi,\eta)= <s(x,y) ~s(x+\xi,y+\eta)>  .
\eeq
Here the angled brackets denote ensemble averages.
For simplicity we have used an isotropic two-dimensional Gaussian
  a.c.f., $\rho(\xi,\eta)=\exp(-[\xi^2+\eta^2]/l^2 )$
where $l$ defines a correlation length. In order to minimise and distinguish
  edge effects we used surfaces which become flat at the outer edge;
  this is not necessary for the method to be applicable.
Studies included the strongly scattering regime of
surfaces with both correlation length and r.m.s. height
of the order of a wavelength.  
With the exception of parallel code mentioned later, all tests were run on
a desktop Pentium 4 3.2GHz machine with 1GB memory running Linux.  

\begin{figure}
\hskip -0.3 true cm
  \includegraphics[height=5.3cm]{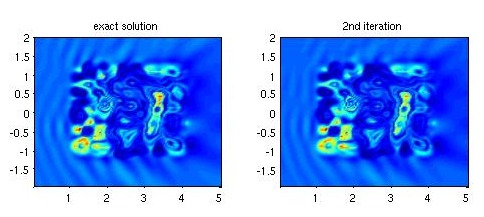}  
\caption{\label{contour} Shaded contour plot of the amplitude of the surface
  fields by (a) exact and (b) iterative solution (2 terms), for
  surface with r.m.s. height and correlation length approximately
  equal to $\lambda$.}
\end{figure}

Comparison was made first against the
full or `exact' inversion of the boundary integral.
The quantity used for the comparison was the surface field. 
Because of the high computational cost of full inversion this
comparison was
carried out for a relatively small surface of $12\times 12$ wavelengths, using
a grid of $120\times 120$ points.  Here the r.m.s. height and correlation
length are approximately equal to $\lambda$.
Contour plots of the amplitude of $\psi$ calculated by
the two methods is shown in Figure \ref{contour}. 
One iteration of the left-right series took around 7 seconds, whereas
``exact'' full inversion took around 23 minutes.   
(The full inversion code at double precision
ran out of memory at this stage so, in this case only, the matrix was
evaluated in single precision.
Iterative code remains in double precision throughout.)

\begin{figure}
  \includegraphics[height=8cm]{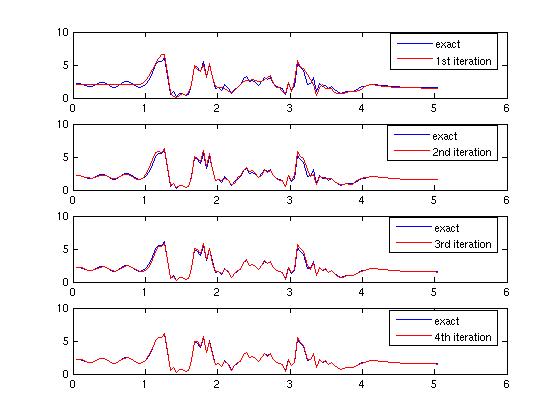} 
\caption{\label{compare} Comparison between exact and successive terms of the
   left-right solution corresponding to Fig. \ref{contour}, along a line
in $x$-direction, for grazing angle $10^o$.}
\end{figure}

\begin{figure}
  \includegraphics[height=8cm]{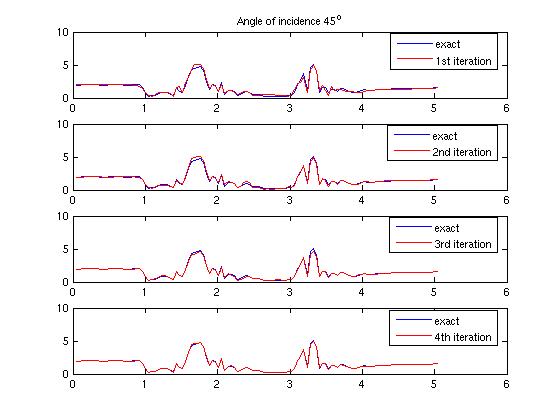} 
\caption{\label{compare2} Comparison for surface as in Figure \ref{compare}, 
for grazing angle of $45^o$.}
\end{figure}

\begin{figure}
  \includegraphics[height=8cm]{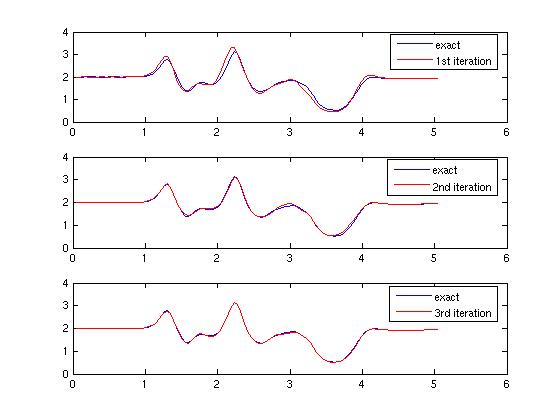} 
\caption{\label{compare3} Comparison between exact and successive terms of the
   left-right solution, for grazing angle $10^o$ due to a smoother
   surface with r.m.s. height $\lambda/20$.}
\end{figure}

In order to illustrate
the convergence, comparison of field values along the mid-line in the
$x$-direction is shown in Figure \ref{compare} 
for the first 4 iterations.
In this case the incident field was a plane wave impinging at an angle
of  10$^o$ from grazing.  Extremely good agreement is found.   
Notice that the oscillatory behaviour at the left is captured at the 2nd
but not the 1st iteration.  (It should be emphasized that
although we found no divergent cases, convergence is not necessarily
guaranteed.  For electromagnetic waves the method
\cite{tran} exhibited divergences apparently due to resonant surface features.)
The solution for an field incident at $45^o$ impinging on 
the same surface is shown in Figure \ref{compare2}, and again
converges rapidly.  
A further comparison (Figure \ref{compare3}) using a `smoother' surface, with
the same correlation lengths but r.m.s. height reduced to 
$\lambda/20$, at $10^o$ from grazing, gives similarly close agreement.

We now consider the application of the code to larger surfaces, in
order further to examine timings and rates of convergence as functions of
incident angle.   Evaluations of the first iterates were carried out
for several cases. As mentioned above, the two main components of the
calculation are a $N$ matrix inversion and a set of Green's
functions evaluation, at each of $N$ range steps.  The matrix
inversion remained a small percentage of the cost in all cases,
and computation time should increase with the square of the number
of unknowns, $M^2N^2$.  The actual computation times were found to
conform closely to this, as shown in Table 1. Times in the second column,
corresponding to the simple optimised integration as described below,
should be regarded as applicable for most surface geometries,
and can easily be reduced further with higher order schemes.


\begin{table}
\begin{center}
\begin{tabular}{| l || r | c |}  

\hline

No. of unknowns & Solution time  & 
Optimised integration \\

\hline \hline

120x120 & 6.9 & 2.6 \\  
\hline
240x160 &48 & 17 \\  
\hline
240x320&198 & 70  \\
\hline
480x320 &774 & 265 \\
\hline
480x480& 1752=29.2min & 605=10.1min \\
\hline
1000x1000 &31870=8.5hrs & 10992 = 3hrs \\

\hline

\end{tabular}
\caption{Computation time on desktop computer}
\end{center}
\end{table}


Note that the algorithm is easily parallelised:
the integration, to which the bulk of computation time is devoted,
can be shared among any number of processors. This has been carried
out using MPI on a Sunfire machine, and as expected the computation speed
increases linearly with the number of processors.
Solution for around $5\times 10^6$ unknowns, on
a waveguide of 550 $\lambda$ in length and 80 $\lambda$
circumference, was obtained in 5.3 hours with standard integration and
under 2 hours using the optimised integration below, on 96 processors.

Strategies are available for reduction of the Green's function
evaluation cost.  One of these is fast multilevel multipole, which can
reduce the time-dependence to $O(NM \log NM)$,
but we found this to have certain disadvantages including relatively high
complexity and memory cost, and accuracy which is not easily regulated.
A much simpler expedient which retains the order of dependence on
the number of unknowns, but reduces the multiplier, is
the following:  
A simple quadrature using all available points was initially used to carry out 
the integration for the left-hand-side of eq. (\ref{abc}).  
The integrand, however, is relatively smooth as a function of
transverse coordinate, and this
increases with spatial separation in $x$.   Thus as the marching
solution proceeds, we can use higher-order integration schemes
utilising far fewer points with little loss of accuracy.
Even a simple trapezium rule, for example, operating on half the 
number of points reduced the computation time by a factor of 3 and
resulted in errors of well under 1\%.
We calculated surface fields on a desktop computer for a surface of 
$48\lambda\times 48\lambda$ (230,000 unknowns) in around 10 minutes, and 
$100\lambda\times 100\lambda$ ($10^6$ unknowns) in 180 minutes.

The same method is applicable to exterior and interior scattering
problems due to various large scatterers and geometries.
Most such geometries involve even better-behaved integrals, and are therefore 
amenable to the above integration strategy. 
Solution for the much larger problem of a waveguide of around 150
wavelengths in length 
and diameter 20 wavelengths (not shown) was calculated on the
desktop computer in around 140 minutes.

\begin{figure}
  \includegraphics[height=5cm]{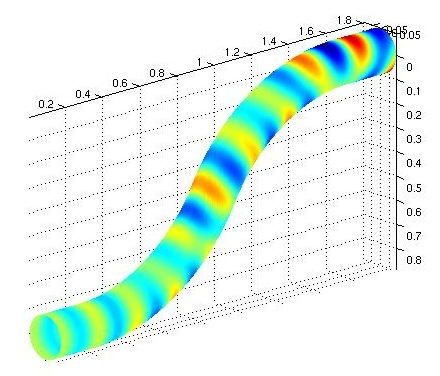} 
  \includegraphics[height=5cm]{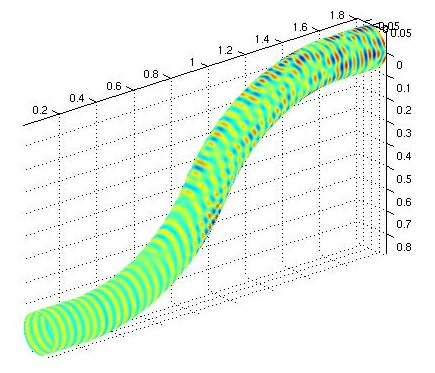} 
\caption{\label{waveguide} Real part of surface field on waveguide 
at two frequencies.}
\end{figure}


\section{Conclusions}

The paper describes the development and application of the left-right 
splitting algorithm for acoustic scattering by rough perfectly
reflecting surfaces and other complex scatterers.   Results have been
validated by comparison with ``exact'' numerical solutions, and
by examining the convergence of the series.    The formulation is
physically-motivated to apply to incident fields at low grazing
angles, although good convergence has been obtained at angles
close to normal incidence.  Problems involving up to $10^6$ unknowns
or more can be solved relatively simply on a standard desktop computer, and
much larger problems still in a few hours on a parallel machine.

The cost of the method scales with the square of the number of
unknowns; this can be improved by application of, say, fast
multipole methods, but this has not been necessary as in this approach the
multiplier is relatively small and can be further reduced by
optimising the integrations.

The terms in the series represent increasing orders of surface
interaction, and this is likely to provide further insight into multiple
scattering mechanisms.


\vskip 2 true cm
\centerline {{\bf Acknowledgements}}
The authors acknowledge partial funding from the DTI eScience programme, and
use of the Cambridge-Cranfield High Performance Computer
Facility.  Many of the ideas arose out of a previous electromagnetic
project supported by BAE Systems and MS is grateful for many helpful
discussions.

\vskip 2 true cm

\vfil\break
\centerline {{\bf Figure Captions}}
\bigbreak
 
\bigbreak
\noindent
Figure \ref{example}: Example rough surface.

\medbreak
\noindent
Figure \ref{schematic}:
Possible paths (a) at 1st iteration, and (b) at 2nd
iteration when reversible paths can occur and add coherently.

\medbreak
\noindent
Figure \ref{contour}: Shaded contour plot of the amplitude of the surface
  fields by (a) exact and (b) iterative solution (2 terms), for
  surface with r.m.s. height and correlation length approximately
  equal to $\lambda$.

\medbreak
\noindent
Figure \ref{compare}:
Comparison between exact and successive terms of the
   left-right solution corresponding to Fig. \ref{contour}, along a line
in $x$-direction, for grazing angle $10^o$.
 
\medbreak
\noindent
Figure \ref{compare2}:
Comparison for surface as in Figure \ref{compare}, 
for grazing angle of $45^o$.

\medbreak
\noindent
Figure \ref{compare3}:
Comparison between exact and successive terms of the
   left-right solution, for grazing angle $10^o$ due to a smoother
   surface with r.m.s. height $\lambda/20$.


\medbreak
\noindent
Figure \ref{waveguide}: 
Real part of surface field on waveguide 
at two frequencies.

\end{document}